\documentclass[aps,prl,superscriptaddress,twocolumn,nobibnotes]{revtex4-1}
\usepackage{graphicx,url,amssymb,amsmath,longtable,rotating,color,units,wasysym,subfigure,epsfig,hyperref,epstopdf}

\begin{document}

\title{Upper Limits on the Stochastic Gravitational-Wave Background from Advanced~LIGO's First Observing Run}


%



\author{%
B.~P.~Abbott,$^{1}$  
R.~Abbott,$^{1}$  
T.~D.~Abbott,$^{2}$  
M.~R.~Abernathy,$^{3}$  
F.~Acernese,$^{4,5}$ 
K.~Ackley,$^{6}$  
C.~Adams,$^{7}$  
T.~Adams,$^{8}$ 
P.~Addesso,$^{9}$  
R.~X.~Adhikari,$^{1}$
% ...
\\
\medskip
(LIGO Scientific Collaboration and Virgo Collaboration) 
\\
\medskip
{{}$^{*}$Deceased, March 2016. }%
}\noaffiliation
\affiliation {LIGO, California Institute of Technology, Pasadena, CA 91125, USA }
\affiliation {Louisiana State University, Baton Rouge, LA 70803, USA }
\affiliation {American University, Washington, D.C. 20016, USA }
\affiliation {Universit\`a di Salerno, Fisciano, I-84084 Salerno, Italy }
\affiliation {INFN, Sezione di Napoli, Complesso Universitario di Monte S.Angelo, I-80126 Napoli, Italy }
\affiliation {University of Florida, Gainesville, FL 32611, USA }
\affiliation {LIGO Livingston Observatory, Livingston, LA 70754, USA }
\affiliation {Laboratoire d'Annecy-le-Vieux de Physique des Particules (LAPP), Universit\'e Savoie Mont Blanc, CNRS/IN2P3, F-74941 Annecy-le-Vieux, France }
\affiliation {University of Sannio at Benevento, I-82100 Benevento, Italy and INFN, Sezione di Napoli, I-80100 Napoli, Italy }
\affiliation {Albert-Einstein-Institut, Max-Planck-Institut f\"ur Gravi\-ta\-tions\-physik, D-30167 Hannover, Germany }
\affiliation {Nikhef, Science Park, 1098 XG Amsterdam, The Netherlands }
\affiliation {LIGO, Massachusetts Institute of Technology, Cambridge, MA 02139, USA }
\affiliation {Instituto Nacional de Pesquisas Espaciais, 12227-010 S\~{a}o Jos\'{e} dos Campos, S\~{a}o Paulo, Brazil }
\affiliation {INFN, Gran Sasso Science Institute, I-67100 L'Aquila, Italy }
\affiliation {INFN, Sezione di Roma Tor Vergata, I-00133 Roma, Italy }
\affiliation {Inter-University Centre for Astronomy and Astrophysics, Pune 411007, India }
\affiliation {International Centre for Theoretical Sciences, Tata Institute of Fundamental Research, Bengaluru 560089, India }
\affiliation {University of Wisconsin-Milwaukee, Milwaukee, WI 53201, USA }
\affiliation {Leibniz Universit\"at Hannover, D-30167 Hannover, Germany }
\affiliation {Universit\`a di Pisa, I-56127 Pisa, Italy }
\affiliation {INFN, Sezione di Pisa, I-56127 Pisa, Italy }
\affiliation {Australian National University, Canberra, Australian Capital Territory 0200, Australia }
\affiliation {California State University Fullerton, Fullerton, CA 92831, USA }
\affiliation {LAL, Univ. Paris-Sud, CNRS/IN2P3, Universit\'e Paris-Saclay, F-91898 Orsay, France }
\affiliation {Chennai Mathematical Institute, Chennai 603103, India }
\affiliation {Universit\`a di Roma Tor Vergata, I-00133 Roma, Italy }
\affiliation {Universit\"at Hamburg, D-22761 Hamburg, Germany }
\affiliation {INFN, Sezione di Roma, I-00185 Roma, Italy }
\affiliation {Albert-Einstein-Institut, Max-Planck-Institut f\"ur Gravitations\-physik, D-14476 Potsdam-Golm, Germany }
\affiliation {APC, AstroParticule et Cosmologie, Universit\'e Paris Diderot, CNRS/IN2P3, CEA/Irfu, Observatoire de Paris, Sorbonne Paris Cit\'e, F-75205 Paris Cedex 13, France }
\affiliation {West Virginia University, Morgantown, WV 26506, USA }
\affiliation {Universit\`a di Perugia, I-06123 Perugia, Italy }
\affiliation {INFN, Sezione di Perugia, I-06123 Perugia, Italy }
\affiliation {European Gravitational Observatory (EGO), I-56021 Cascina, Pisa, Italy }
\affiliation {Syracuse University, Syracuse, NY 13244, USA }
\affiliation {SUPA, University of Glasgow, Glasgow G12 8QQ, United Kingdom }
\affiliation {LIGO Hanford Observatory, Richland, WA 99352, USA }
\affiliation {Wigner RCP, RMKI, H-1121 Budapest, Konkoly Thege Mikl\'os \'ut 29-33, Hungary }
\affiliation {Columbia University, New York, NY 10027, USA }
\affiliation {Stanford University, Stanford, CA 94305, USA }
\affiliation {Universit\`a di Padova, Dipartimento di Fisica e Astronomia, I-35131 Padova, Italy }
\affiliation {INFN, Sezione di Padova, I-35131 Padova, Italy }
\affiliation {Nicolaus Copernicus Astronomical Center, Polish Academy of Sciences, 00-716, Warsaw, Poland }
\affiliation {Center for Relativistic Astrophysics and School of Physics, Georgia Institute of Technology, Atlanta, GA 30332, USA }
\affiliation {University of Birmingham, Birmingham B15 2TT, United Kingdom }
\affiliation {Universit\`a degli Studi di Genova, I-16146 Genova, Italy }
\affiliation {INFN, Sezione di Genova, I-16146 Genova, Italy }
\affiliation {RRCAT, Indore MP 452013, India }
\affiliation {Faculty of Physics, Lomonosov Moscow State University, Moscow 119991, Russia }
\affiliation {SUPA, University of the West of Scotland, Paisley PA1 2BE, United Kingdom }
\affiliation {Caltech CaRT, Pasadena, CA 91125, USA }
\affiliation {University of Western Australia, Crawley, Western Australia 6009, Australia }
\affiliation {Department of Astrophysics/IMAPP, Radboud University Nijmegen, P.O. Box 9010, 6500 GL Nijmegen, The Netherlands }
\affiliation {Artemis, Universit\'e C\^ote d'Azur, CNRS, Observatoire C\^ote d'Azur, CS 34229, F-06304 Nice Cedex 4, France }
\affiliation {Institut de Physique de Rennes, CNRS, Universit\'e de Rennes 1, F-35042 Rennes, France }
\affiliation {Washington State University, Pullman, WA 99164, USA }
\affiliation {Universit\`a degli Studi di Urbino 'Carlo Bo', I-61029 Urbino, Italy }
\affiliation {INFN, Sezione di Firenze, I-50019 Sesto Fiorentino, Firenze, Italy }
\affiliation {University of Oregon, Eugene, OR 97403, USA }
\affiliation {Laboratoire Kastler Brossel, UPMC-Sorbonne Universit\'es, CNRS, ENS-PSL Research University, Coll\`ege de France, F-75005 Paris, France }
\affiliation {Carleton College, Northfield, MN 55057, USA }
\affiliation {Astronomical Observatory Warsaw University, 00-478 Warsaw, Poland }
\affiliation {VU University Amsterdam, 1081 HV Amsterdam, The Netherlands }
\affiliation {University of Maryland, College Park, MD 20742, USA }
\affiliation {Laboratoire des Mat\'eriaux Avanc\'es (LMA), CNRS/IN2P3, F-69622 Villeurbanne, France }
\affiliation {Universit\'e Claude Bernard Lyon 1, F-69622 Villeurbanne, France }
\affiliation {Universit\`a di Napoli 'Federico II', Complesso Universitario di Monte S.Angelo, I-80126 Napoli, Italy }
\affiliation {NASA/Goddard Space Flight Center, Greenbelt, MD 20771, USA }
\affiliation {RESCEU, University of Tokyo, Tokyo, 113-0033, Japan. }
\affiliation {University of Adelaide, Adelaide, South Australia 5005, Australia }
\affiliation {Tsinghua University, Beijing 100084, China }
\affiliation {Texas Tech University, Lubbock, TX 79409, USA }
\affiliation {The University of Mississippi, University, MS 38677, USA }
\affiliation {The Pennsylvania State University, University Park, PA 16802, USA }
\affiliation {National Tsing Hua University, Hsinchu City, 30013 Taiwan, Republic of China }
\affiliation {Charles Sturt University, Wagga Wagga, New South Wales 2678, Australia }
\affiliation {University of Chicago, Chicago, IL 60637, USA }
\affiliation {Kenyon College, Gambier, OH 43022, USA }
\affiliation {Korea Institute of Science and Technology Information, Daejeon 305-806, Korea }
\affiliation {University of Cambridge, Cambridge CB2 1TN, United Kingdom }
\affiliation {Universit\`a di Roma 'La Sapienza', I-00185 Roma, Italy }
\affiliation {University of Brussels, Brussels 1050, Belgium }
\affiliation {Sonoma State University, Rohnert Park, CA 94928, USA }
\affiliation {Montana State University, Bozeman, MT 59717, USA }
\affiliation {Center for Interdisciplinary Exploration \& Research in Astrophysics (CIERA), Northwestern University, Evanston, IL 60208, USA }
\affiliation {Universitat de les Illes Balears, IAC3---IEEC, E-07122 Palma de Mallorca, Spain }
\affiliation {The University of Texas Rio Grande Valley, Brownsville, TX 78520, USA }
\affiliation {Bellevue College, Bellevue, WA 98007, USA }
\affiliation {Institute for Plasma Research, Bhat, Gandhinagar 382428, India }
\affiliation {The University of Sheffield, Sheffield S10 2TN, United Kingdom }
\affiliation {California State University, Los Angeles, 5154 State University Dr, Los Angeles, CA 90032, USA }
\affiliation {Universit\`a di Trento, Dipartimento di Fisica, I-38123 Povo, Trento, Italy }
\affiliation {INFN, Trento Institute for Fundamental Physics and Applications, I-38123 Povo, Trento, Italy }
\affiliation {Cardiff University, Cardiff CF24 3AA, United Kingdom }
\affiliation {Montclair State University, Montclair, NJ 07043, USA }
\affiliation {National Astronomical Observatory of Japan, 2-21-1 Osawa, Mitaka, Tokyo 181-8588, Japan }
\affiliation {Canadian Institute for Theoretical Astrophysics, University of Toronto, Toronto, Ontario M5S 3H8, Canada }
\affiliation {MTA E\"otv\"os University, ``Lendulet'' Astrophysics Research Group, Budapest 1117, Hungary }
\affiliation {School of Mathematics, University of Edinburgh, Edinburgh EH9 3FD, United Kingdom }
\affiliation {University and Institute of Advanced Research, Gandhinagar, Gujarat 382007, India }
\affiliation {IISER-TVM, CET Campus, Trivandrum Kerala 695016, India }
\affiliation {University of Szeged, D\'om t\'er 9, Szeged 6720, Hungary }
\affiliation {Embry-Riddle Aeronautical University, Prescott, AZ 86301, USA }
\affiliation {Tata Institute of Fundamental Research, Mumbai 400005, India }
\affiliation {INAF, Osservatorio Astronomico di Capodimonte, I-80131, Napoli, Italy }
\affiliation {University of Michigan, Ann Arbor, MI 48109, USA }
\affiliation {Rochester Institute of Technology, Rochester, NY 14623, USA }
\affiliation {NCSA, University of Illinois at Urbana-Champaign, Urbana, IL 61801, USA }
\affiliation {University of Bia{\l }ystok, 15-424 Bia{\l }ystok, Poland }
\affiliation {SUPA, University of Strathclyde, Glasgow G1 1XQ, United Kingdom }
\affiliation {University of Southampton, Southampton SO17 1BJ, United Kingdom }
\affiliation {University of Washington Bothell, 18115 Campus Way NE, Bothell, WA 98011, USA }
\affiliation {Institute of Applied Physics, Nizhny Novgorod, 603950, Russia }
\affiliation {Seoul National University, Seoul 151-742, Korea }
\affiliation {Inje University Gimhae, 621-749 South Gyeongsang, Korea }
\affiliation {National Institute for Mathematical Sciences, Daejeon 305-390, Korea }
\affiliation {Pusan National University, Busan 609-735, Korea }
\affiliation {NCBJ, 05-400 \'Swierk-Otwock, Poland }
\affiliation {Institute of Mathematics, Polish Academy of Sciences, 00656 Warsaw, Poland }
\affiliation {The School of Physics \& Astronomy, Monash University, Clayton 3800, Victoria, Australia }
\affiliation {Hanyang University, Seoul 133-791, Korea }
\affiliation {The Chinese University of Hong Kong, Shatin, NT, Hong Kong }
\affiliation {University of Alabama in Huntsville, Huntsville, AL 35899, USA }
\affiliation {ESPCI, CNRS, F-75005 Paris, France }
\affiliation {University of Minnesota, Minneapolis, MN 55455, USA }
\affiliation {Universit\`a di Camerino, Dipartimento di Fisica, I-62032 Camerino, Italy }
\affiliation {Southern University and A\&M College, Baton Rouge, LA 70813, USA }
\affiliation {The University of Melbourne, Parkville, Victoria 3010, Australia }
\affiliation {College of William and Mary, Williamsburg, VA 23187, USA }
\affiliation {Instituto de F\'\i sica Te\'orica, University Estadual Paulista/ICTP South American Institute for Fundamental Research, S\~ao Paulo SP 01140-070, Brazil }
\affiliation {Whitman College, 345 Boyer Avenue, Walla Walla, WA 99362 USA }
\affiliation {Universit\'e de Lyon, F-69361 Lyon, France }
\affiliation {Hobart and William Smith Colleges, Geneva, NY 14456, USA }
\affiliation {Janusz Gil Institute of Astronomy, University of Zielona G\'ora, 65-265 Zielona G\'ora, Poland }
\affiliation {King's College London, University of London, London WC2R 2LS, United Kingdom }
\affiliation {IISER-Kolkata, Mohanpur, West Bengal 741252, India }
\affiliation {Indian Institute of Technology, Gandhinagar Ahmedabad Gujarat 382424, India }
\affiliation {Andrews University, Berrien Springs, MI 49104, USA }
\affiliation {Universit\`a di Siena, I-53100 Siena, Italy }
\affiliation {Trinity University, San Antonio, TX 78212, USA }
\affiliation {University of Washington, Seattle, WA 98195, USA }
\affiliation {Abilene Christian University, Abilene, TX 79699, USA }



\noaffiliation

\begin{abstract}
A wide variety of astrophysical and cosmological sources are expected to contribute to a stochastic gravitational-wave background. Following the observations of GW150914 and GW151226, the rate and mass of coalescing binary black holes appear to be greater than many previous expectations. As a result, the stochastic background from unresolved compact binary coalescences is expected to be particularly loud. We perform a search for the isotropic stochastic gravitational$\mbox{-}$wave background using data from Advanced LIGO's first observing run.  The data display no evidence of a stochastic gravitational$\mbox{-}$wave signal. We constrain the dimensionless energy density of gravitational waves to be $\Omega_0<1.7\times 10^{-7}$ with $95\%$ confidence, assuming a flat energy density spectrum in the most sensitive part of the LIGO band $(20-86\ {\rm Hz})$. This is a factor of $\sim$33 times more sensitive than previous measurements. We also constrain arbitrary power-law spectra. Finally, we investigate the implications of this search for the background of binary black holes using an astrophysical model for the background.
\end{abstract}

\maketitle

\emph{Introduction.}--- Many astrophysical and cosmological phenomena are expected to contribute to a stochastic gravitational$\mbox{-}$wave background, henceforth, simply refered to as a ``background''.  These include unresolved compact binary coalescences of both black holes and neutron stars \citep{2013MNRAS.431..882Z,PhysRevD.84.124037,PhysRevD.85.104024,PhysRevD.84.084004,2011ApJ...739...86Z}, rotating neutron stars \citep{PhysRevD.87.063004,2012PhRvD..86j4007R,2011ApJ...729...59Z}, supernovae \citep{2009MNRAS.398..293M,2010MNRAS.409L.132Z,PhysRevD.72.084001,PhysRevD.73.104024}, cosmic strings \citep{2005PhRvD..71f3510D,1976JPhA....9.1387K,2002PhLB..536..185S,2007PhRvL..98k1101S}, inflationary models \citep{1994PhRvD..50.1157B,1979JETPL..30..682S,2007PhRvL..99v1301E,2012PhRvD..85b3525B,2012PhRvD..85b3534C,2013arXiv1305.5855L,1997PhRvD..55..435T,2006JCAP...04..010E}, phase transitions \cite{Kosowsky1992rz,Kamionkowski:1993fg,thumb}, and the pre-Big Bang scenario \cite{Gasperini:1992em,Gasperini:1993hu,PBBpaper,Gasperini:2016gre}. 
The variety of mechanisms potentially contributing to the background provides the opportunity to study a number of different environments within the Universe.

The recent detections of binary black hole (BBH) coalescences by Advanced~LIGO \citep{PhysRevLett.116.061102,PhysRevLett.116.241103} suggest that the Universe may be rich with coalescing BBHs. 
While events like GW150914 and GW151226 are loud enough to be clearly detected, we expect there to be many more events that are too far away to be individually resolved and that contribute to the background. Since this BBH population originates from sources that are too distant to be individually detected, the stochastic search probes a distinct population of binaries compared to nearby sources \cite{sgwbPE}.  The background from these binaries provides complementary information to individually resolved binary coalescences \citep{PhysRevLett.116.131102}.

In this Letter, we report on the search for an isotropic background using data from Advanced~LIGO's first observing run O1. We search for the background by cross-correlating data streams from the two separate LIGO detectors and looking for a coherent signal.  We find no evidence for the background and place the best upper limits to date on the energy density of the background in the LIGO frequency band.  We also update the implications for a BBH background using all the data from O1.


\emph{Data.}---Before this analysis, the best limits on the background from Initial LIGO and Virgo data were obtained using 2009--2010 \citep{2014PhRvL.113w1101A} and 2005--2007 data \citep{Aasi:2014jkh}. In this work we use data from the upgraded Advanced~LIGO observatories in Hanford, WA (H1) and Livingston, LA (L1) \citep{cqg.32.074001.15}. We analyze O1 data from September 18, 2015 15:00 UTC-January 12, 2016 16:00 UTC.


\emph{Method.}--- We define the background energy density spectrum as \cite{1999PhRvD..59j2001A}
\begin{equation}
    \Omega_{\textnormal{GW}}(f) = \frac{f}{\rho_c}\frac{d\rho_{\textnormal{GW}}}{df},
    \label{eq:omega_gw_definition}
\end{equation}
where $f$ is the frequency, $\rho_c=3c^2 H_0^2/(8\pi G)$ is the critical energy density to close the Universe (numerically, $\rho_c=7.8\times10^{-9}~{\rm erg/cm^3}$ using the Hubble constant $H_0=68\ {\rm km\ s^{-1}\ Mpc^{-1}}$ from~\citep{Ade:2015xua,Grieb:2016uuo}), and $d \rho_{\textnormal{GW}}$ is the gravitational$\mbox{-}$wave energy density in the frequency range from $f$ to $f+df$. For the LIGO frequency band, most theoretical models for $\Omega_{\textnormal{GW}}(f)$ can be approximated as a power law in frequency~\citep{1999PhRvD..59j2001A,2007ApJ...659..918A,2012PhRvD..85l2001A}:
\begin{equation}
    \Omega_{\textnormal{GW}}(f) = \Omega_{\alpha}\left(\frac{f}{f_{\textnormal{ref}}}\right)^{\alpha}.
    \label{eq:omega_gw_powerlaw_def}
\end{equation}
Following~\cite{PhysRevLett.116.131102}, we assume a reference frequency of $25\ {\rm Hz}$, which corresponds to the most sensitive band of the LIGO stochastic search for a detector network operating at design sensitivity.
The variable $\Omega_{\alpha}$ characterizes the background amplitude across the sensitive frequency band.
Past analyses have used $\alpha=0$ and $\alpha=3$ to represent cosmologically and astrophysically motivated background models respectively~\citep{2007ApJ...659..918A,2009Natur.460..990A,2011PhRvL.107A1102A,2012PhRvD..85l2001A,2014PhRvL.113w1101A}.
In this analysis we use these two spectral indices but also include limits on the background spectrum assuming $\alpha=2/3$, which describes the background dominated by compact binary inspirals \cite{PhysRevLett.116.131102,Meacher:2015iua}. This choice of spectral index is especially interesting given the loud background from BBHs inferred from recent Advanced LIGO detections in O1 \citep{PhysRevLett.116.061102,PhysRevLett.116.241103,PhysRevLett.116.131102,TheLIGOScientific:2016pea}.

Our search uses a cross-correlation method optimized to search for the background using the pair of LIGO detectors~\cite{1999PhRvD..59j2001A}. As discussed for instance in \cite{Romano:2016dpx}, cross-correlation is preferred to auto-correlation methods because the noise variances in each detector are not known sufficiently well to perform subtraction of the noise auto-power. We define the estimator
\begin{equation}
\hat{Y}_\alpha=\int_{-\infty}^{\infty}df\,\int_{-\infty}^{\infty}df'\,\delta_T(f-f'){\tilde s}^{*}_{1}(f){\tilde s}_{2}(f'){\tilde Q}_\alpha(f')
\label{eq:cc_estimator}
\end{equation}
with variance
\begin{equation}
\sigma^2_Y{\approx}\frac{T}{2}\int_{0}^{\infty}df\,P_1(f)P_2(f)|{{\tilde Q}_\alpha(f)}|^2,
\label{eq:theor_sigma}
\end{equation}
where $\tilde s_{1,2}(f)$ are the Fourier transforms of the strain time series data from the two detectors, $\delta_T(f-f')$ is a finite-time approximation to the Dirac delta function, $T$ is the observation time, $P_{1,2}$ are the one-sided power spectral densities for the detectors, and $\tilde Q_\alpha(f)$ is a filter function to optimize the search \footnote{The Hubble constant appears explicitly, rather than being absorbed into $\lambda_\alpha$, to emphasize that the estimator for $\Omega_{\rm GW}$ depends on the measured value of $H_0$.},
\begin{equation}
{\tilde Q}_\alpha(f)={\lambda_\alpha}\frac{{\gamma}(f)H_0^2}{f^{3}P_1(f)P_2(f)}\left(\frac{f}{f_{\rm ref}}\right)^\alpha.
\label{eq:optimal_filter}
\end{equation}
The spatial separation and relative orientation of the two detectors are accounted for in the overlap reduction function, $\gamma(f)$~\cite{PhysRevD.46.5250} and the normalization constant $\lambda_\alpha$ is chosen such that $\langle \hat Y_\alpha \rangle = \Omega_\alpha$.


\emph{Data Quality.}---For this analysis, the strain time series data are down$\mbox{-}$sampled to 4096~Hz from 16384~Hz and separated into 50\%-overlapping 192~s segments, as in \cite{2007ApJ...659..918A}.  The segments are Hann$\mbox{-}$windowed and high$\mbox{-}$pass filtered with a $16^{\rm th}$ order Butterworth digital filter with knee frequency of 11~Hz. The data are coarse-grained to a frequency resolution of $0.031$~Hz. This is a finer frequency resolution than was used in previous analyses due to the need to remove many finely spaced lines at low frequencies.

We apply cuts in the time and frequency domains, following \cite{2014PhRvL.113w1101A}. The total live time after all time domain vetoes have been applied was 29.85 days. These cuts remove 35\% of the time-series data. The frequency domain cuts remove 21\% of the observing band. In the Supplementary Matrial \cite{technical_supplement}, which includes Refs.~\cite{0264-9381-33-13-134001,hardware-inj,schumann,wsubtract}, we discuss in more detail the removed times and frequencies, the recovery of hardware and software injections, and an analysis of correlated noise due to geophysical Schumann resonances. 


\emph{Results.}---Our search finds no evidence of the background, and the data are consistent with statistical fluctuations, assuming Gaussian noise. The integrand of Equation~\ref{eq:cc_estimator}, multiplied by $df=0.031\ {\rm Hz}$, gives an estimator for $\Omega_0$ in each frequency bin. We plot this quantity, along with $\pm 2\sigma$ error bars, in Figure~\ref{fig_1}. To check for Gaussianity, we employ a noise model that the estimator in each frequency bin is drawn from a Gaussian distribution with zero mean with the standard deviation of that frequency bin. We obtain a $\chi^2$ per degree of freedom of 0.92, indicating that the data are consistent with Gaussian noise. 

Consequently, we are able to place upper bounds on the energy density present in the background. For $\alpha=0$, we place the bound $\Omega_0<1.7\times10^{-7}$ at $95\%$ confidence, where $99\%$ of the sensitivity comes in the frequency band $20-86\ {\rm Hz}$. This is a factor of 33 times more sensitive than the previous best limit at these frequencies~\cite{2014PhRvL.113w1101A}.

\begin{figure} 
\includegraphics[width=0.5\textwidth]{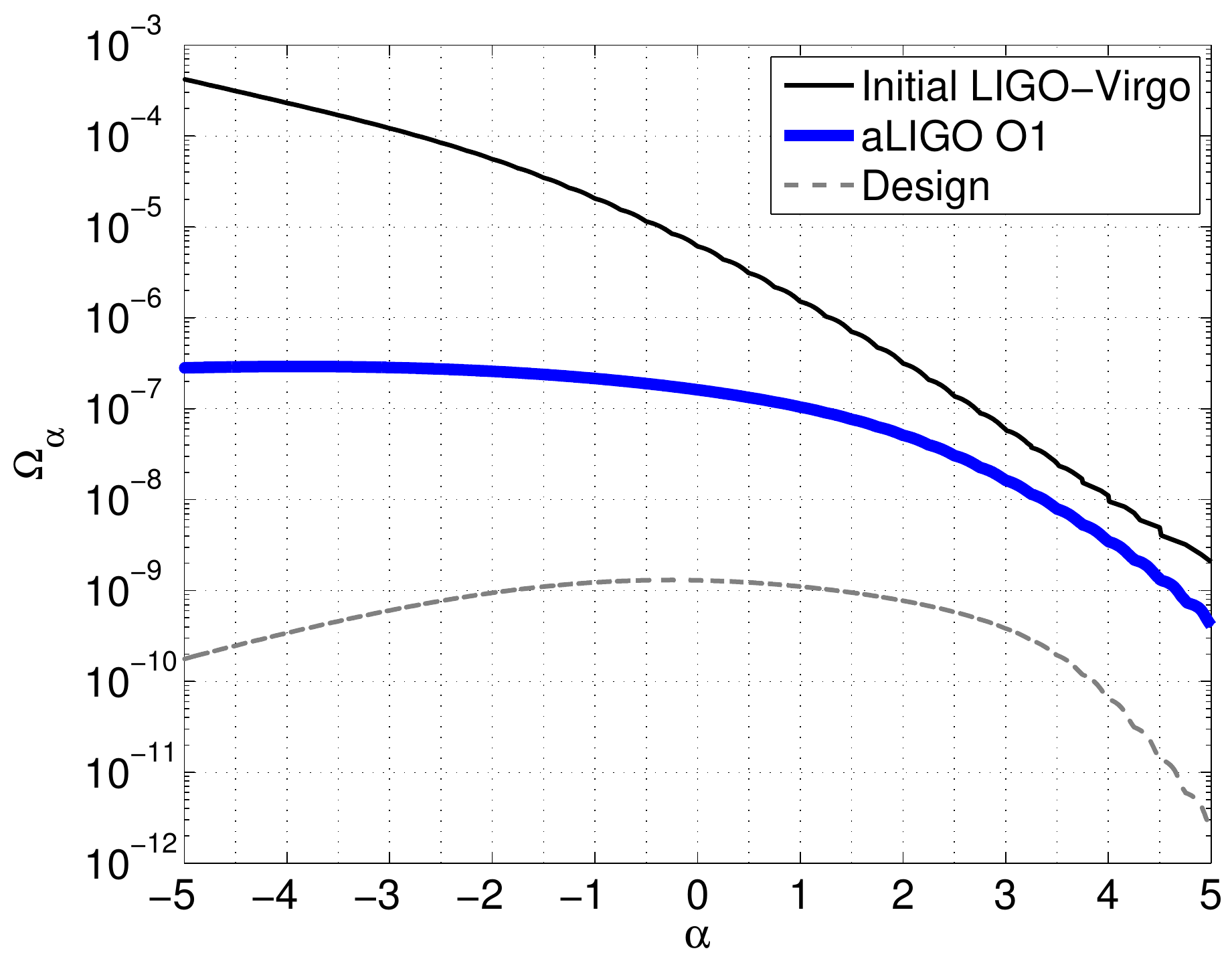}
\caption{Following \cite{stoch_paramest}, we present 95 \% confidence contours in the $\Omega_\alpha-\alpha$ plane. The region above these curves is excluded at $95\%$ confidence. We show the constraints coming from the final science run of Initial LIGO-Virgo \cite{2014PhRvL.113w1101A} and from O1 data. Finally, we display the projected (not observed) design sensitivity to $\Omega_\alpha$ and $\alpha$ for Advanced LIGO and Virgo \cite{Aasi:2013wya}.}
\label{fig_2}
\end{figure}

Following \cite{stoch_paramest}, we show $95\%$ confidence contours in the $\Omega_\alpha-\alpha$ plane in Figure~\ref{fig_2} by computing the joint posterior for $\Omega_\alpha$ and $\alpha$. In addition, in Table~\ref{table_1}, we report upper limits on the energy density for specific fixed values of the spectral index, marginalizing over amplitude calibration uncertainty \citep{2014JPHCS.484a2027W} using the conservative estimates of $11.8\%$ for H1 and $13.4\%$ for L1. Phase calibration uncertainties are negligible.

\begin{figure} 
\includegraphics[width=0.5\textwidth]{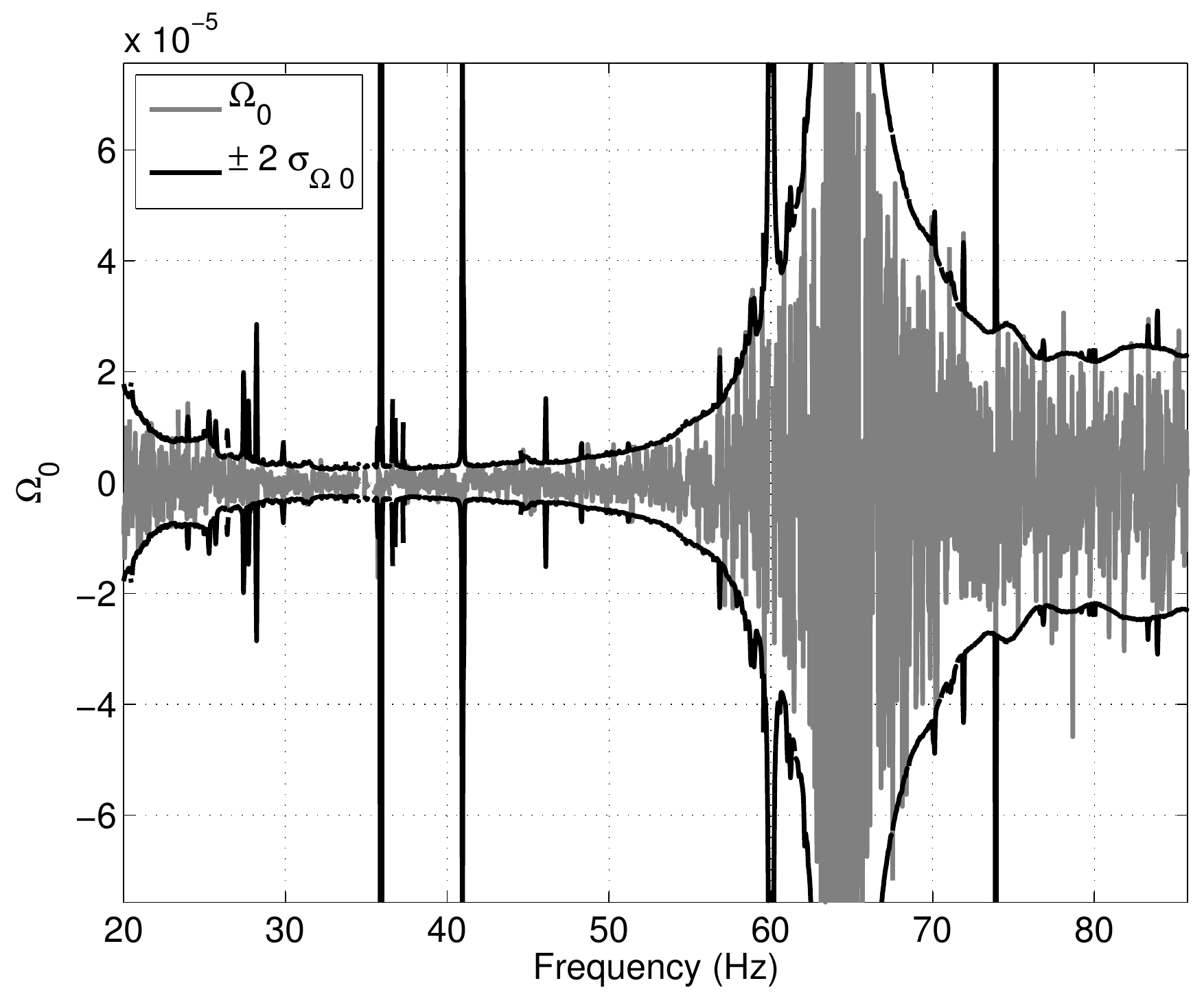}
\caption{We show the estimator for $\Omega_0$ in each frequency bin, along with $\pm 2 \sigma$ error bars, in the frequency band that contains 99\% of the sensitivity for $\alpha=0$. The loss of sensitivity at around 65 Hz is due to a zero in the overlap reduction function. There are several lines associated with known instrumental artifacts which do not lead to excess cross-correlation. The data are consistent with Gaussian noise, as described in the Results section.}
\label{fig_1}
\end{figure}

\begin{table*} 
\begin{tabular}{|c|c|c|c|c|}
\hline
Spectral index $\alpha$ & Frequency band with $99\%$ sensitivity  & Amplitude $\Omega_{\alpha}$ & 95\% CL upper limit & Previous limits \cite{2014PhRvL.113w1101A} \\
\hline
\hline
\hline
 0 & $20-85.8\ {\rm Hz}$ & $(4.4\pm5.9)\times10^{-8}$ & $1.7\times10^{-7}$ & $5.6\times10^{-6}$ \\
\hline
$2/3 $ & $20-98.2\ {\rm Hz}$ & $(3.5\pm4.4)\times10^{-8}$ & $1.3\times10^{-7}$ & -- \\
\hline
3 & $20-305\ {\rm Hz}$ & $(3.7\pm6.5)\times10^{-9}$ &  $1.7\times10^{-8}$ & $7.6 \times 10^{-8}$ \\
\hline
\end{tabular}
\caption{The frequency band with 99\% of the sensitivity are shown, along with the point estimate and standard deviation for the amplitude of the background, and $95\%$ confidence level upper limits using O1 data for three values of the spectral index, $\alpha=0,2/3,3$. We also show the previous upper limits using Initial LIGO-Virgo data. \label{tsbl-1}}
\label{table_1}
\end{table*}

We also compare our results with the limits placed at high frequencies from the two co-located detectors at the Hanford site (H1 and H2). In \cite{Aasi:2014jkh}, the limit $\Omega_3<7.7\times 10^{-4}$ in the frequency band $460-1000\ {\rm Hz}$ was obtained for the spectral index $\alpha=3$ and $f_{\rm ref}=900\ {\rm Hz}$. Using this same frequency band, and using the cross-correlated data between the Hanford and Livingston detectors, we place a limit $\Omega_3<1.7\times 10^{-2}$ for $f_{\rm ref}=900\ {\rm Hz}$. This is about a factor of 22 larger than the limit from the co-located detectors, in part due to the loss in sensitivity of a stochastic search from cross-correlating detectors at different spatial locations.

\begin{figure*} 
\includegraphics[width=\textwidth]{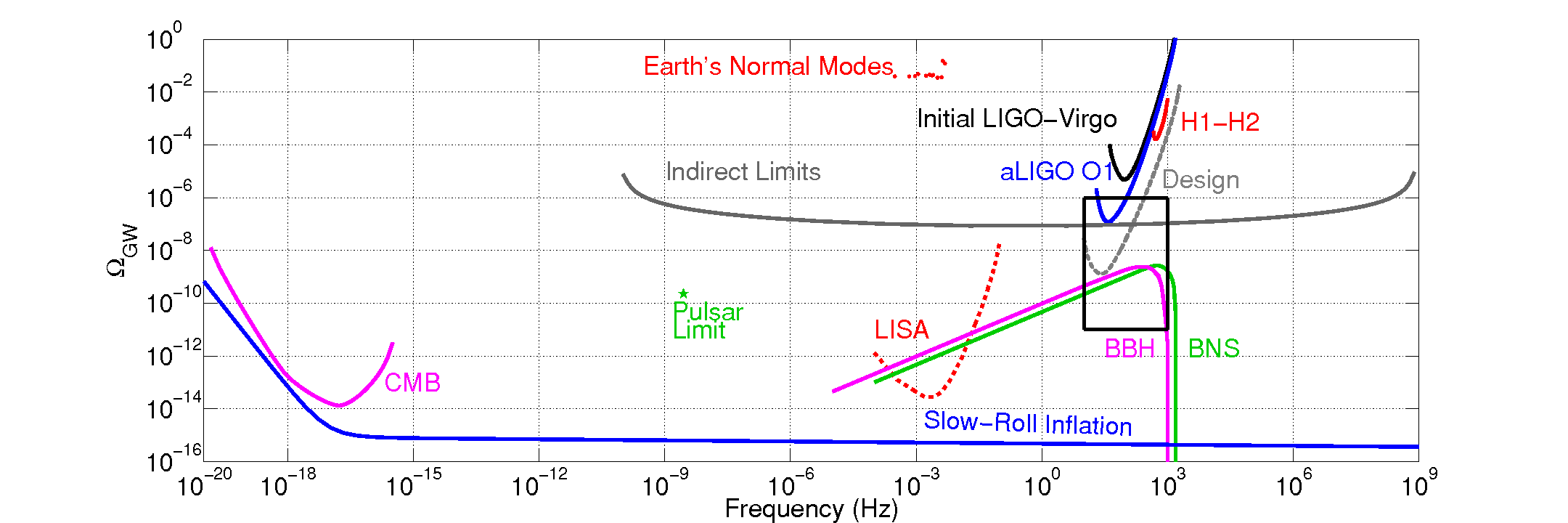}
\caption{Presented here are constraints on the background in PI form~\cite{locus}, as well as some representative models, across many decades in frequency. We compare the limits from ground-based interferometers from the final science run of Initial LIGO-Virgo, the co-located detectors at Hanford (H1-H2), Advanced LIGO (aLIGO) O1, and the projected design sensitivity of the advanced detector network assuming two years of coincident data, with constraints from other measurements: CMB measurements at low multipole moments \cite{Ade:2011ah}, indirect limits from the Cosmic Microwave Background (CMB) and Big-Bang Nucleosynthesis \cite{Pagano:2015hma,Lasky:2015lej}, pulsar timing \cite{Lasky:2015lej}, and from the ringing of Earth's normal modes \cite{Coughlin:2014xua}.  We also show projected limits from a space-based detector such as LISA \cite{Cornish:2001qi,Crowder:2005nr,locus}, following the assumptions of \cite{locus}. We extend the BNS and BBH distributions using an $f^{2/3}$ power-law down to low frequencies, with a low-frequency cut-off imposed where the inspiral time-scale is of order the Hubble scale. In Figure \ref{fig_5}, we show the region in the black box in more detail.}
\label{fig_3}
\end{figure*}

In Figure~\ref{fig_3}, we show the constraints from this analysis and from previous analyses using other detectors, theoretical predictions, and the expected sensitivity of future measurements by LIGO-Virgo and by the Laser Interferometer Space Antenna (LISA). Where applicable, we show constraints using power-law integrated curves (PI curves) \cite{locus}, which account for the broadband nature of the search by integrating a range of power-law signals over the sensitive frequency band of the detector. By construction, any power-law spectrum which crosses a PI curve is detectable with ${\rm SNR}\geq 2$. 

The blue curve labeled `aLIGO O1' in Figure~\ref{fig_3} shows the measured O1 PI curve. We also display the PI curve for the final science run of Initial LIGO and Virgo \cite{2014PhRvL.113w1101A}, H1-H2 \cite{Aasi:2014jkh}, as well as the projected design sensitivity for the advanced detector network. The curve labeled `Design' assumes 2 years of co-incident data taken with both Advanced LIGO and Virgo operating at design sensitivity, using the projections in \cite{Aasi:2013wya}. For the sake of comparison, the measured O1 PI curve at $\alpha=0$ is 1.6 times larger than the projected PI curve at $\alpha=0$ using the projections in \cite{Aasi:2013wya} and 29.85 days of live time, which is fairly good agreement between predicted and achieved sensitivity. Finally, in red we present the projected sensitivity of a space-based detector with similar sensitivity to LISA, using the PI curve presented in \cite{locus} computed using the projections in \cite{Cornish:2001qi,Crowder:2005nr}.

We compare these constraints with direct limits from the ringing of Earth's normal modes \cite{Coughlin:2014xua}, indirect limits from the Cosmic Microwave Background (CMB) and Big Bang Nucleosynthesis (BBN) \cite{Pagano:2015hma}, and limits from pulsar timing arrays \cite{Lasky:2015lej} and CMB measurements at low multipole moments \cite{Ade:2011ah}.

In addition, we give examples of several models which can contribute to the background. We show the background expected from slow-roll inflation with a tensor-to-scalar-ratio $r=0.11$ (the upper limit allowed by Planck \cite{Ade:2015xua}). We also show examples of the BBH coalescence model, and the binary neutron star (BNS) coalescence model, which we describe below. As noted in \cite{2016PhRvL.116w1102S}, LISA is likely to be able to detect the BBH background of the size considered here.


\emph{Astrophysical Implications.}---In order to model the background from binary systems we will follow the approach of \cite{PhysRevLett.116.131102}. We divide the compact binary population into classes labeled by $k$ \cite{Kim2003,Abbott:2016nhf}. Each class has distinct values of source parameters (for example the masses), which we denote by $\theta_k$. The total astrophysical background is a sum over the contributions in each class. The contribution of class $k$ to the background may be written in terms of an integral over the redshift $z$ as \cite{2011RAA....11..369R,2011ApJ...739...86Z,2011PhRvD..84h4004R,2011PhRvD..84l4037M,2012PhRvD..85j4024W,2013PhRvD..87d2002W,2013MNRAS.431..882Z,2015A&A...574A..58K}
\begin{equation}
\Omega_{\rm GW}(f;\theta_k)=\frac{f}{\rho_c H_0} \int_0^{z_{\rm max}} dz \frac{R_m(z;\theta_k)\frac{dE_{\rm GW}}{df}(f_s;\theta_k)}{(1+z) E(\Omega_M,\Omega_\Lambda,z)},  
\end{equation}
where $R_m(z;\theta_k)$ is the binary merger rate per unit comoving volume per unit time, $dE_{\rm GW}/df(f_s,\theta_k)$ is the energy spectrum emitted by a single binary evaluated in terms of the source frequency $f_s=(1+z)f$, and $E(\Omega_M,\Omega_\Lambda,z)=\sqrt{\Omega_M (1+z)^3+\Omega_\Lambda}$ accounts for the dependence of comoving volume on cosmology. We use cosmological parameters from Planck \cite{Ade:2015xua}, and $\Omega_M=1-\Omega_\Lambda=0.308$. 

The energy spectrum $dE_{\rm GW}/df$ is determined from the strain waveform of the binary system. The dominant contribution to the background comes from the inspiral phase, however for BBH we include the merger and ringdown phases using the waveforms from \cite{2008PhRvD..77j4017A,2011ApJ...739...86Z} with the modifications from \cite{Ajith:2011}. We choose to cut off the redshift integral at $z_{\rm max}=10$. Redshifts larger than five contribute little to the integral due to the small number of stars formed at such high redshift \cite{2011RAA....11..369R,2011ApJ...739...86Z,2011PhRvD..84h4004R,2011PhRvD..84l4037M,2012PhRvD..85j4024W,2013PhRvD..87d2002W,2013MNRAS.431..882Z,2015A&A...574A..58K,sgwbPE}.

To compute the binary merger rate $R_m(z;\theta_k)$, we use the same assumptions as in \cite{PhysRevLett.116.131102}, unless stated otherwise. For the BNS case, we assume that the minimal time between the formation and the coalescence of the binary is $t_{\rm min}=20\ {\rm Myr}$, following for instance~\cite{Meacher:2015iua}. This is to be compared to $t_{\rm min}=50\ {\rm Myr}$ for BBH~\cite{dominik,PhysRevLett.116.131102}.

As was emphasized in \cite{gw150914astro}, heavy stellar mass black holes are expected to form in regions of low metallicity, which are associated with weaker stellar winds. To account for this effect, following \cite{PhysRevLett.116.131102}, for binary systems with chirp masses larger than $30 \ M_\odot$, we use only the fraction of stars that form in an environment with metallicity $Z<Z_\odot/2$. For BBH (and BNS) systems with smaller masses, we do not use a cutoff. However, we note that it makes little difference whether or not the cutoff is applied to high masses.

\begin{figure}[tb] 
\includegraphics[width=0.5\textwidth]{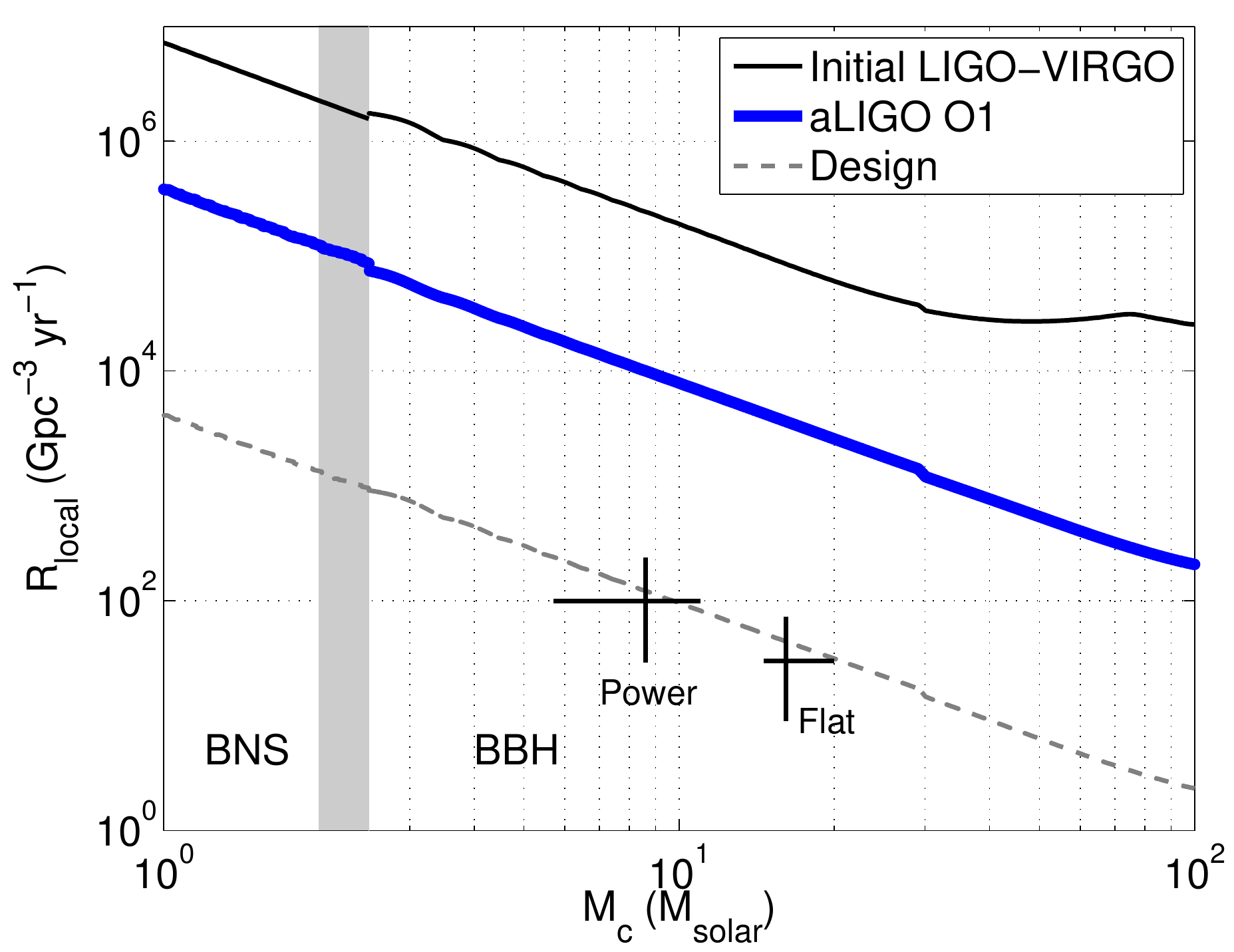}
\centering
\caption{Displayed here are the 95\% confidence contours on the local rate and average chirp mass parameters, using the model described in the Astrophysical Implications section. In addition to the constraint from Advanced LIGO (aLIGO) O1 data, we show the constraint from the final science run of Initial LIGO-Virgo, and the projected design sensitivity of Advanced LIGO-Virgo. We also show the median rate with 90\% uncertainty inferred from O1 data for the power-law and flat-log mass distributions \cite{TheLIGOScientific:2016pea}, along with the band containing 68\% of the chirp mass for each distribution. The gray band separates BNS from BBH backgrounds. The dip at 30 $M_\odot$ is due to the metallicity cutoff, as described in the Astrophysical Implications section.} 
\label{fig_4}  

\end{figure}

With the model defined above, the free parameters are the local merger rate $R_{\rm local}=R_m(0;\theta_k)$ and the average chirp mass $M_c$. The distribution of the chirp mass has little effect on the spectrum for a fixed average chirp mass \cite{2011ApJ...739...86Z}. 

We place upper limits at 95\% confidence in the $M_c-R_{\rm local}$ plane, which are shown in Figure~\ref{fig_4}. Alongside the O1 results, we show the limits using Initial LIGO-Virgo data, as well as projected sensitivity of the advanced detector network. The limits presented here are about 10 times more sensitive than those placed with Initial LIGO-Virgo data. Furthermore, the future runs of the advanced detectors are expected to yield another factor of 100 improvement in sensitivity in $R_{\rm local}$ for a given average chirp mass. We also show the local rate and chirp mass inferred from direct detections of BBH mergers during O1 \cite{Abbott:2016nhf,TheLIGOScientific:2016pea}. Comparing the projected design sensitivity on $R_{\rm local}$ and $M_c$, with the values inferred from BBH observations in O1, suggests that it may be possible for the advanced detector network to detect the astrophysical BBH background.

\begin{figure}[ht] 
\vspace{0.1cm}
\includegraphics[width=0.5\textwidth]{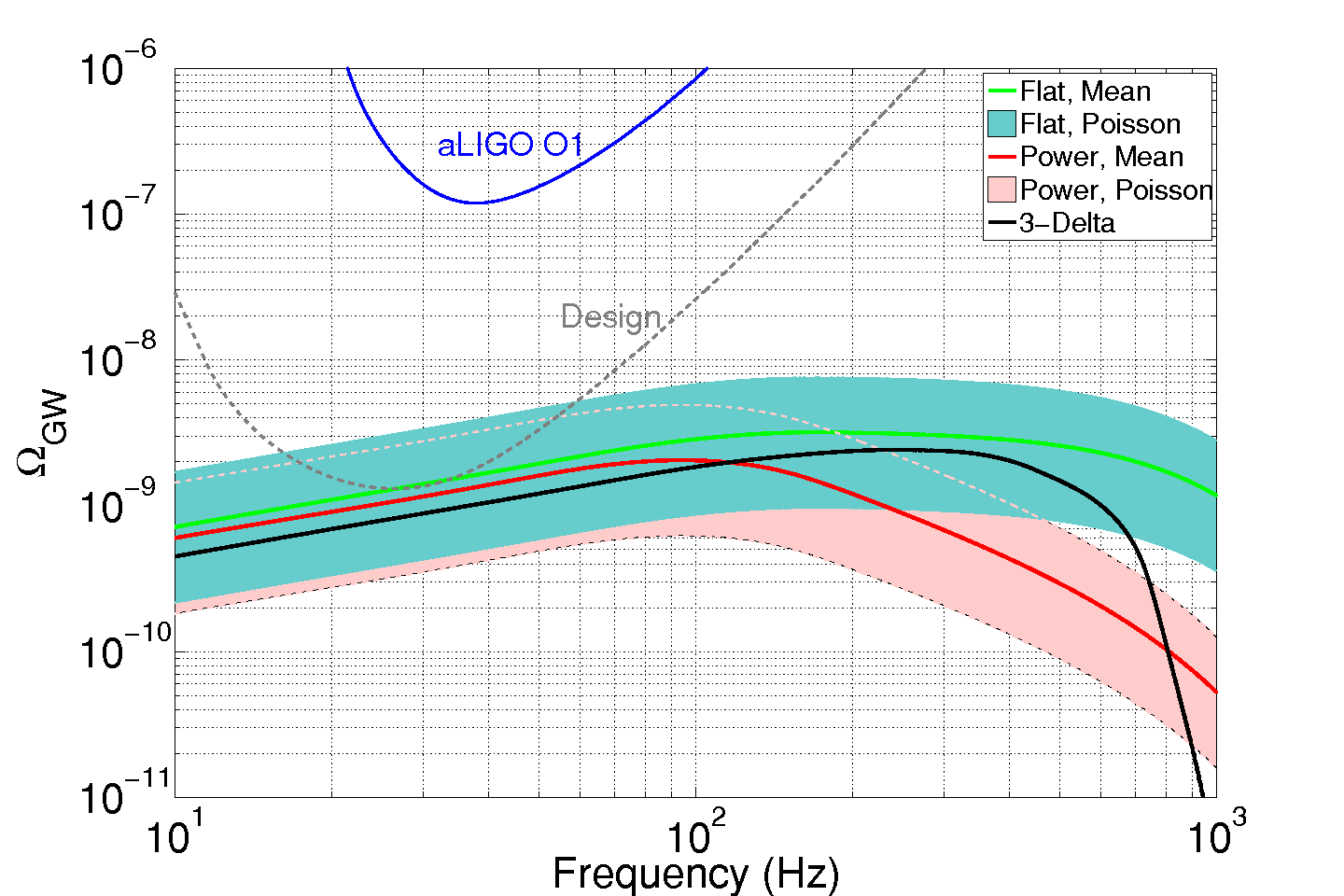}
\caption{We present a range of potential spectra for a BBH background, using the flat-log, power-law, and 3-delta mass distribution models described in \cite{gw150914astro,TheLIGOScientific:2016pea}, with the local rate inferred from the O1 detections \cite{TheLIGOScientific:2016pea}. For the flat-log and power-law distributions, we show the 90\% Poisson uncertainty band due to the uncertainty in the local rate measurement. In addition, we show the measured O1 PI curve and the projected PI curve for Advanced LIGO-Virgo operating at design sensitivity.}
\label{fig_5}
\end{figure}

Finally, instead of treating the chirp mass and local merger rate as free parameters, we can use the information from individually observed BBHs to compute the corresponding background, see Figure~\ref{fig_5}. To do this, we use the same model described above and we adopt the three rate models described in \cite{TheLIGOScientific:2016pea}. Specifically, we consider the three-events-based, power-law, and flat-log distributions of component masses. In each case, the rate at redshift $z=0$ is normalized to the local rate derived from the O1 detections. With these assumptions we compute the background, including statistical uncertainty bands showing the $90\%$ uncertainty in the local rate. The three rate models agree well in the sensitive frequency band of advanced detectors (10-100 Hz). Note also that the final sensitivity of the advanced detectors may be sufficient to detect this background.


\emph{Conclusions.}---The results presented here represent the first search for the stochastic gravitational-wave background made with the Advanced~LIGO detectors.  With no evidence of a stochastic signal, we place an upper limit of $\Omega_0 < 1.7 \times 10^{-7}$ on the GW energy density, for a spectral index $\alpha=0$. This is $\sim$33 times more sensitive than previous direct measurements in this frequency band.  We also constrain the binary coalescence parameters of chirp mass and local merger rate. For fixed chirp mass below the high mass threshold of $30\ M_\odot$, the constraint on the merger rate is improved by a factor of $\sim 24$, while for fixed merger rate, the constraint on the chirp mass is improved by a factor of $\sim 7$, as can be seen from Figure~\ref{fig_4}. Finally, we update the background predictions due to BBH coalescences using data from O1. In this work we have focused the implications of our results for an astrophysical BBH background, as this provides the most promising candidate for first detecting the background. The implications of our search for other astrophysical and cosmological models can be seen in Figure~\ref{fig_3}. There is also an upcoming publication that will study implications for cosmic string models in more detail.

These O1 results are a glimpse of the improvements in sensitivity to be seen in upcoming years. As the advanced detectors reach design sensitivity, there is a reasonable possibility of detecting the background due to BBHs. Even if no detection is made with these future searches, the searches will be able to constrain important cosmological and astrophysical background models.

\emph{Acknowledgments.}--- The authors gratefully acknowledge the support of the United States
National Science Foundation (NSF) for the construction and operation of the
LIGO Laboratory and Advanced LIGO as well as the Science and Technology Facilities Council (STFC) of the
United Kingdom, the Max-Planck-Society (MPS), and the State of
Niedersachsen/Germany for support of the construction of Advanced LIGO 
and construction and operation of the GEO600 detector. 
Additional support for Advanced LIGO was provided by the Australian Research Council.
The authors gratefully acknowledge the Italian Istituto Nazionale di Fisica Nucleare (INFN),  
the French Centre National de la Recherche Scientifique (CNRS) and
the Foundation for Fundamental Research on Matter supported by the Netherlands Organisation for Scientific Research, 
for the construction and operation of the Virgo detector
and the creation and support  of the EGO consortium. 
The authors also gratefully acknowledge research support from these agencies as well as by 
the Council of Scientific and Industrial Research of India, 
Department of Science and Technology, India,
Science \& Engineering Research Board (SERB), India,
Ministry of Human Resource Development, India,
the Spanish Ministerio de Econom\'ia y Competitividad,
the Conselleria d'Economia i Competitivitat and Conselleria d'Educaci\'o, Cultura i Universitats of the Govern de les Illes Balears,
the National Science Centre of Poland,
the European Commission,
the Royal Society, 
the Scottish Funding Council, 
the Scottish Universities Physics Alliance, 
the Hungarian Scientific Research Fund (OTKA),
the Lyon Institute of Origins (LIO),
the National Research Foundation of Korea,
Industry Canada and the Province of Ontario through the Ministry of Economic Development and Innovation, 
the Natural Science and Engineering Research Council Canada,
Canadian Institute for Advanced Research,
the Brazilian Ministry of Science, Technology, and Innovation,
Funda\c{c}\~ao de Amparo \`a Pesquisa do Estado de S\~ao Paulo (FAPESP),
Russian Foundation for Basic Research,
the Leverhulme Trust, 
the Research Corporation, 
Ministry of Science and Technology (MOST), Taiwan
and
the Kavli Foundation.
The authors gratefully acknowledge the support of the NSF, STFC, MPS, INFN, CNRS and the
State of Niedersachsen/Germany for provision of computational resources.
This article has been assigned the document number LIGO-P1600258-v19.

\bibliography{iso_bibliography_v3}

\newpage
\newpage

\section*{Supplement--Upper Limits on the Stochastic Gravitational-Wave Background from Advanced~LIGO's First Observing Run}

\maketitle

In this supplement we describe in more detail how the data in the main text are analyzed. Data used in the analysis are from times when both detectors are in a low-noise observing mode.  We exclude certain times and frequencies based on auxiliary channels that established them as instrumental effects within the detectors.
 
We remove times due to known instrumental artifacts, such as radio frequency (RF) glitching and electronics saturations~\cite{0264-9381-33-13-134001}, or due to simulated signals (referred to as hardware injections) generated by coherently moving the interferometer mirrors \cite{hardware-inj}. We also exclude segments associated with detections of gravitational waves.  
Data are also excluded when the detectors' noise power spectra vary by more than 20\% over the course of three 192s segments. This cut is performed to remove non-stationary noise, and has been used in previous analyses \citep{2014PhRvL.113w1101A}. A dedicated study has verified that removing variations of 20\% provides a close-to-optimal balance between the false positive and false negative rates. The total live time with all vetoes applied, for 192s segments, is  29.85 days. These cuts remove $35\%$ of the time-series data.

We exclude frequencies known to be associated with instrumental artifacts, such as vibrations of the test mass suspensions and calibration lines.  We also remove frequencies that are known to be instrumentally correlated between the two LIGO detectors.  As an example, we detected a comb-like structure (a series of lines evenly spaced in frequency) at half Hz frequencies with 1 Hz separation. This structure was coherent between the two sites and subsequently observed in auxiliary channels. The contributing frequency bins were not included in the analysis. The frequency domain cuts remove 21\% of the observing band within each segment.

To verify the data analysis cuts described above, we introduce an artificial time shift of 1~s between the two sites. This effectively blinds the analysis by removing correlations due to a broadband gravitational-wave signal, while maintaining instrumental correlations with coherence times greater than 1~s. This method also allows us to identify additional instrumental artifacts that are not identified using the cuts above, without biasing our analysis of the data. Upon studying the time-shifted data with the analysis cuts described above, we find no excess correlation, which is consistent with statistical expectations of uncorrelated Gaussian noise.

\begin{figure}[b] 
\includegraphics[width=0.5\textwidth]{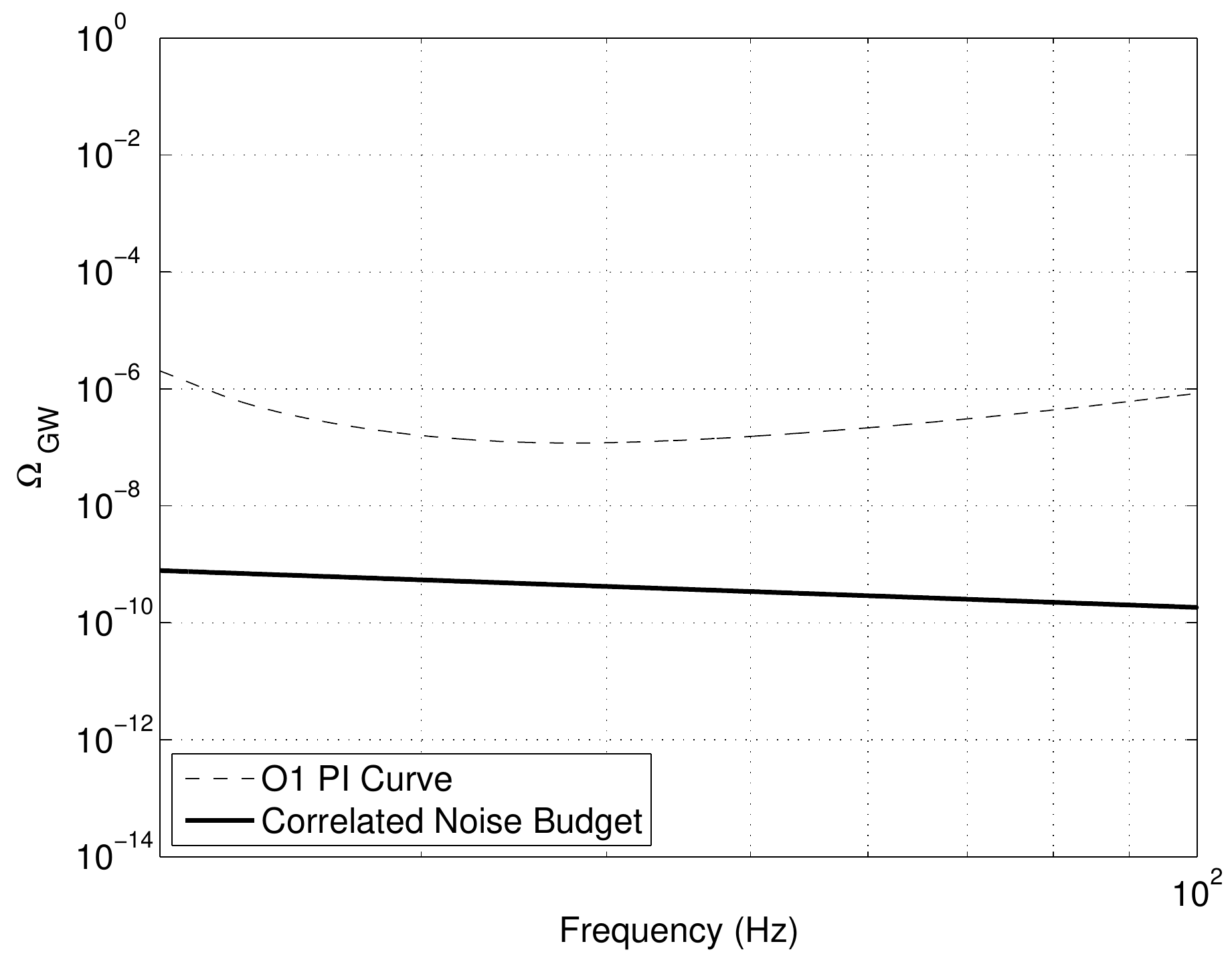}
\caption{We show the O1 power-law integrated curve (PI curve) along with the correlated noise budget as described in the text. The noise budget falling below the O1 PI curve indicates that correlated noise does not affect the O1 analysis.}
\label{fig_1}
\end{figure}

As a test of the detectors and the analysis pipeline, we simulate a strong stochastic signal both by a hardware injection and by a software injection (made by adding a coherent signal to the data streams).  The injected background signals were isotropic and Gaussian, with an amplitude of $\Omega_0=8.7\times10^{-5}$ and a duration of 600~s.  Both types of injections were successfully recovered within 1 $\sigma$ uncertainty: the hardware injection measured $(8.8\pm 0.6)\times10^{-5}$ and the software injection measured $(9.0\pm0.6)\times10^{-5}$. 

Finally, we study the possibility of correlated noise between H1 and L1 so that we may be confident that the systematic error in our measurements is negligible. After accounting for narrowband correlation detector artifacts arising from digital systems, we estimate the contamination from the environment. Previous investigations have identified geophysical Schumann resonances as the most likely source of correlated environmental noise~\cite{schumann,wsubtract}. Excitations in the spherical shell cavity formed between the surface of the Earth and the ionosphere cause magnetic fields to be correlated over great distances, comparable to the separation between H1 and L1. The magnetic fields, in turn, can couple mechanically to the test mass through the suspension system or electronically~\cite{wsubtract}. In order to ascertain the systematic error from environmental correlated noise, we construct a correlated noise budget. We employ a number of conservative assumptions in order to estimate the worst-case-scenario contamination.

The first step is to measure the frequency-dependent coupling of the detector to ambient magnetic fields using external coils as an actuator~\cite{schumann,wsubtract}. It is not practical to induce fields that act on the entire detector simultaneously, so we measure the coupling at each test mass. Next, we use magnetometers to measure the magnetic coherence between the two sites. Using the method described in \cite{schumann,wsubtract}, we combine the magnetic cross-power spectra and the coupling functions to estimate the worst-case correlated noise from Schumann resonances $\Omega_\text{noise}(f)$. Our conservative noise budget for O1 corresponds to the solid black curve in Figure~\ref{fig_1}. This curve is obtained by fitting a power law to the magnetic noise budget. We compare the noise budget to the power-law integrated energy density spectrum (dashed black curve)~\cite{locus}, which represent the statistical uncertainty of the stochastic search. During O1, the correlated noise is sufficiently low as to be ignored, contributing much less than one sigma. (If the correlated noise estimate was significant, the noise budget would be comparable to or in excess of the dashed curve in the region of $\sim$20-30 Hz.) Work is ongoing to monitor and mitigate correlated noise for future.

\end{document}